# Development of an eReaxFF Force Field for BZY20 Solid Oxide Electrocatalysis

Md Jamil Hossain[1,2,#], Prashik Gaikwad[2], Yun Kyung Shin[1], Jessica Schulze[1], Kate Penrod[1], Meng Li[2], Yuxiao Lin[2], Gorakh Pawar[2] and Adri C. T. van Duin[1,*]

[1]*Department of Mechanical Engineering, The Pennsylvania State University, University Park, Pennsylvania 16802, USA*

[2]*Energy and Environment Science & Technology Directorate, Idaho National Laboratory, Idaho Falls, Idaho 83402, USA*

[#]*Currently at School of Engineering, Brown University, Providence, Rhode Island 02912, USA*

## ABSTRACT

Electrocatalysis is a catalytic process where the rate of an electrochemical reaction occurring at the electrode-electrolyte interface can be controlled by varying the electrical potential. Electrocatalysis can be applied to generate hydrogen which can be stored for future use in fuel cells for clean electricity. The use of solid oxide in electrocatalysis specially in hydrogen evolution reaction is promising. However, further improvements are essential in order to meet the ever-increasing global energy demand. Improvement of the performance of these high energy chemical systems is directly linked to the understanding and improving the complex physical and chemical phenomena and exchanges that take place at their different interfaces. To enable large length and time scale atomistic simulations of solid oxide electrocatalysis for hydrogen generation, we developed an eReaxFF force field for barium zirconate doped with 20 mol% of yttrium (BZY20). All parameters for the eReaxFF were optimized to reproduce quantum mechanical (QM) calculations on relevant condensed phase and cluster systems describing oxygen vacancies, vacancy migrations, water adsorption, water splitting and hydrogen generation on the surfaces of the BZY20 solid oxide. Using the developed force field, we performed zero-voltage molecular dynamics simulations to observe water adsorption and the eventual hydrogen production. Based on our simulation results, we conclude that this force field sets a stage for the introduction of explicit electron concept in order to simulate electron conductivity, electron leakage and non-zero-voltage effects on hydrogen generation. Overall, we demonstrate how atomistic-scale simulations can enhance our understanding of processes at interfaces in solid oxide materials.

## Introduction

Sustainable energy is essential to human societies. Currently, the major source of energy is fossil fuels. As world population continues to grow and the amount of fossil fuels begins to diminish, it may not be possible to provide the amount of energy demanded by the human population by only using fossil fuels to generate energy. Moreover, energy extraction from carbon-burning has a detrimental impact on the environment. The depletion of fossil fuels necessitates alternate and clean energy sources. As the availability of renewable electricity increases, electrochemical devices, such as fuel cells and electrolysis devices, can help facilitate conversion between electrical and chemical forms, allowing for storage of electrical energy in chemical form, or electricity-driven conversions to generate useful chemical compounds.

Electrocatalysis involves a catalytic process where bonds break and form for oxidation and reduction reactions through the direct transfer of electrons and ions. The electrocatalysts must process the ability to lower the overpotential of these reactions[1] as well as to adsorb key intermediate species as it provides alternate energy pathways[2]. Electrocatalytic processes take place in fuel cells, electrolysis devices or even in electrochemical energy storage in batteries, such as metal–air batteries[3].

One of the key areas in electrocatalysis is clean energy conversion – an electrochemical conversion processes that convert molecules in the atmosphere (e.g., water, carbon dioxide, and nitrogen) into important fuels and chemicals (e.g., hydrogen, hydrocarbons, oxygenates and ammonia)[4]. These fuels can be used in fuel cells to produce clean electricity. This current study will focus on hydrogen production. Decarbonized hydrogen production pathways based on renewable energy sources (i.e., wind energy, photovoltaic energy, hydro energy) and water electrolysis are considered attractive and promising solutions for a sustainable future[5,6]. The water electrolysis process consists of employing electricity generated from renewable energy sources to split pure water into oxygen and hydrogen[7]. The water-splitting reaction, which consists of the hydrogen evolution reaction (HER; $2H^+ + 2e^- \rightarrow H_2$), is quite noteworthy in the production of hydrogen[8,9]. The generated and stored hydrogen can be employed for different applications such as transportation, energy storage, power-to-gas, and industry[5,10]. Details of hydrogen production, storage, applications and catalysts utilized are reported in the review article[11].

The reaction rate of an electrocatalyst system can be improved by increasing the number of active sites on a given electrode e.g., through increased loading or improved catalyst structuring to expose more active sites to facilitate binding different reaction intermediates and transition states in different ways. This can be done by alloying, doping, or the introduction of defects.

Electrocatalysts can be composed of materials ranging from metals, metal oxides, carbonaceous materials etc[12]. Across the wide operating range of electrochemical devices, electrolytes can range from aqueous ionic solutions, ions dispersed in organic solvents, ionic liquids, ionic polymers, molten salts, solid acids, solid bases, oxygen-conducting solid oxides such as yttria-stabilized zirconia (YSZ), and oxygen and proton conducting solid oxides such as yttrium-doped barium zirconate (BZY). Transition metal oxides (TMOs) have been widely used in electrocatalysis because of their cost-effective, stable, accessible, and environmentally benign characteristics. Currently, TMOs are one of the most promising classes of electrocatalysts for HER. Barium zirconate doped with 20 mol% of yttrium (BZY20) is an almost pure proton conductor in the presence of water vapor or/and hydrogen containing atmospheres below 650 °C. BZY20 has a high proton conductivity with a small activation energy[13]. The BZY20 electrolyte is also chemically very stable[14]. However, its current density efficiency suffers due to electronic leakage[15].

Despite a significant amount of experimental effort, detailed mechanisms of the intricate interfacial physical and chemical phenomena in solid oxide electrolysis devices have not been explored comprehensively. The atomistic simulations can effectively enlighten the issues pertinent to the electrolysis device chemistry and the change in the interface properties during operation. Modeling techniques can help understand the complexities of catalyst and electrode/electrolyte interface in the atomistic- and molecular-level while providing insights in the depictions of the solvent, cations, and anions near the interface, as well as the mechanisms and reaction barriers of key steps involving proton/ electron transfers in any reaction environment – lenient to harsh. Modeling the Electrocatalytic processes require to describe bond-breaking and bond-making processes which make a quantum chemical treatment such as density functional theory

(DFT) useful and various application of DFT are detailed in the review article[12]. However, the inability of DFT to simulate large length and timescale make them limited in this application. On the other hand, empirical force field-based methods such as reactive force fields can simulate reactions at larger length and time scales.

The ReaxFF reactive force field method has been employed in many chemical systems including solid oxide fuel cells[16]. However, ReaxFF method has its limitations as far electron conductivity is concerned. The eReaxFF method[17] is an extension of the ReaxFF method with a description of an explicit pseudo-classical electron-like particle. This method has been applied to simulate electronic motion in carbon-based-systems[17], lithium-electrolyte systems[18] and was recently extended to Ag metal systems[19] and polyethylene electrical breakdown[20]. No other empirical force field-based methods have been reported in the literature for the description of solid oxide electrocatalysis probably due to the complexity of the different elements and their interactions in the solid oxide and also electrocatalysis itself.

In this manuscript, we describe the development and application of an eReaxFF force field for BZY20 solid oxide electrocatalysis to study oxygen vacancy migrations, water adsorption, water splitting and hydrogen generation at different surfaces. First, we generated a suite of relevant condensed phased density functional theory (DFT) data describing BZY20 bulk and slab structures with and without the presence of oxygen vacancies and water adsorption energies, water splitting and hydrogen generation energies at different surfaces (100, 110) with different surface terminations. Next, these DFT data were subsequently added to existing ReaxFF training data for YSZ solid oxide[16]. Then, we trained the eReaxFF Ba/Zr/Y/O atom, bond, angle and dihedral parameters with the aim to find the optimal reproduction of the DFT data. Finally, we demonstrate the quality of the newly developed force field in simulating the water adsorption, water splitting and eventual proton transfer and hydrogen generation at zero-voltage of the BZY20 solid oxide. Our quality of the force field trends opens up an avenue for introducing the explicit electron component to the force field for the purpose of simulating non-zero voltage affects and identifying causes for electronic leakage.

## Methods

**Density functional theory (DFT)**

We performed a series of periodic DFT calculations on the bulk and several slab models with different surface terminations of BZY20 solid oxide material. For our periodic DFT calculations we used a combination of VASP and CP2K planewave method using PBE functionals. We have considered different configurations for each case based on the pristine structures as well as structures with oxygen vacancies at different locations and concentrations. Next, we calculated slab models with different concentrations of hydrogenation. Based on these DFT calculations, we obtained oxygen migration energies, water adsorption energies, hydrogen generation energies and subsequently put all these data in an eReaxFF force field training set. We have also added the training set data of the previously published YSZ solid oxide force field[16] to our training set.

**eReaxFF**

eReaxFF method is an extension of the standard ReaxFF method which incorporates a pseudo-classical explicit electron scheme. The general expression of the eReaxFF energy is

$$E_{system} = E_{bond} + E_{over} + E_{under} + E_{lp} + E_{val} + E_{tor} + E_{vdWaals} + E_{coulomb} + E_{nucl-elec} + E_{elec}$$

where partial energy contributions include bond, over-coordination penalty and under-coordination stability, lone pair, valence, torsion, non-bonded van der Waals, Coulomb, electron-nucleus interactions and electron-electron interactions respectively. All the many-body bonded interaction and nonbonded interaction terms of ReaxFF are retained with modifications made to the over-coordination penalty, under-coordination stability and lone pair terms to incorporate explicit electrons. In addition, new energy functionals are included to account for electron-nucleus and electron-electron interactions.

In ReaxFF, the bonded and non-bonded interactions are calculated independently, that is, the valency and number of lone pair electrons of an atom type are independent of atomic charge. Thus, the bond orders are fully de-coupled from the charges - as such, a hydrogen ion ($H^+$) is considered capable of forming bonds. eReaxFF, on the other hand, fixes this shortcoming by introducing variable valency and number of lone-pair electrons of an atom depending on the proximity of explicit electrons. The bond orders and over-coordination, under-coordination and lone pair energies are calculated based on the corrected valence electrons. Removing an electron from an atom decreases the atom's valency and consequently increases the over-coordination of an existing bond, resulting in a larger over-coordination energy penalty, and therefore reduces the bond order associated with that atom and weakens the bond. In contrast to ReaxFF, this scheme has enabled the eReaxFF method to counteract the formation of unphysical bonds such as $H^+$–$H^+$ due to the higher over-coordination penalty stemming from the loss of valence electron from the H atom. To compute atomic charges, eReaxFF uses the atom-condensed Kohn–Sham density functional theory approximated to the second order (ACKS2) charge calculation scheme instead of the electronegativity equalization method (EEM) or the charge equilibrium method (QEq) used by most ReaxFF force fields. In the current eReaxFF force fields, the mass of the electron particles is considered 1 amu in order to allow femto-second MD timesteps.

## Force field training

The eReaxFF force field parameters were developed and trained with an aim to reproduce quantum chemistry-based density functional theory (DFT) data. Parameter optimization was conducted using a successive one-parameter search technique[21] to minimize the following expression for the error:

$$error = \sum_{i=1}^{n} \left( \frac{x_{i,lit} - x_{i,eReaxFF}}{\sigma_i} \right)^2 \qquad (6.1)$$

where $x_{i,lit}$ and $x_{i,eReaxFF}$ are the target quantum chemistry/experimental data and eReaxFF values of the i$^{th}$ entry of the force field training set and $\sigma_i$ is the inverse weight that determines how accurately that particular data needs to be fitted with the QC. The summation of all errors in the training set provides the overall error, which is one of the measures of the quality of the force field.

We have taken BZY20 bulk structure with three different Y doping locations. From Figure 1, it can be seen that the Y doping of the 3rd kind is the most energetically favorable one according to both eReaxFF and DFT. The rest of the analysis will be based on the structure corresponding to the 3rd kind of doping.

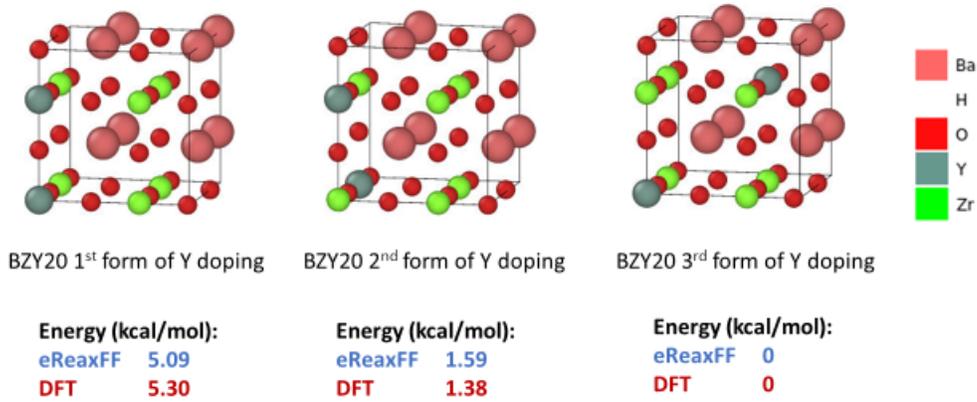

**Fig. 1** | BZY20 bulk structures with different Y doping position and comparison of their energies calculated by both eReaxFF and DFT.

Now, oxygen vacancies are created by removing one or to oxygen atoms at different locations and shown in Figure 2a. The 't1 O vac' and 't2 O vac' structures have one oxygen vacancy while '2 t1 O vac' and '2 t2 O vac' structures have two oxygen vacancies. The corresponding eReaxFF and DFT energies are plotted in the bar graph in Figure 2b. The negative number indicates that the existence of oxygen vacancies in the structures are not energetically favorable. Similar qualitative trends between eReaxFF and DFT energies can be seen.

(a)                                              (b)

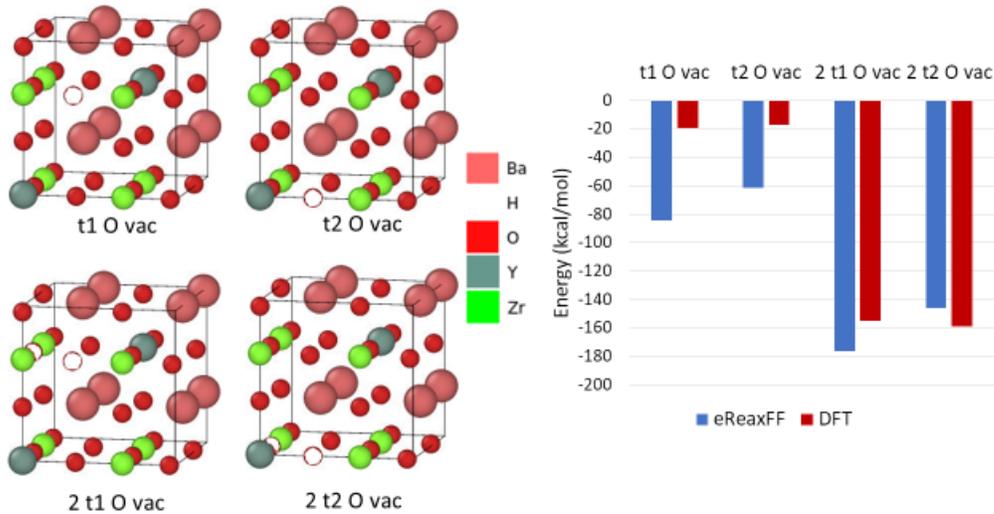

**Fig. 2 |** (a) BZY20 3rd form of Y doping with different number and sites of oxygen vacancies (marked in white circles) and (b) comparison of their energies calculated by both eReaxFF and DFT.

Figure 3 shows the equation of state of BZY20 bulk structure calculated using both eReaxFF and DFT. Both methods are in excellent agreement with each other, especially around the equilibrium volume.

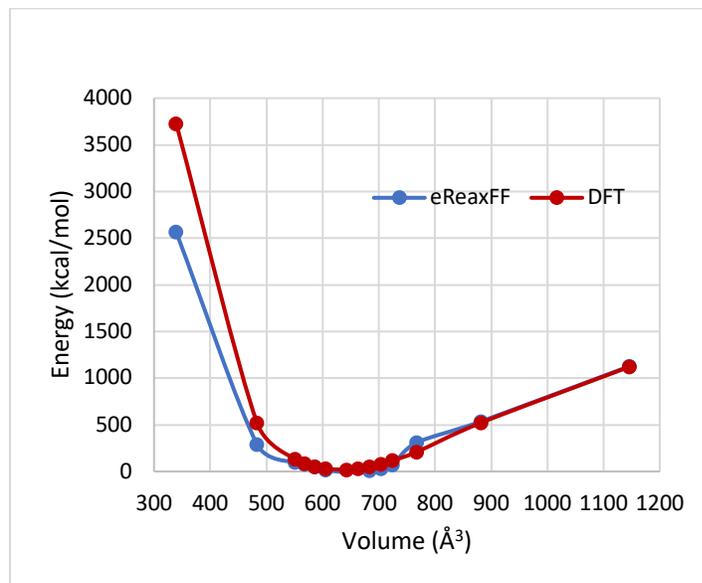

**Fig. 3 |** (a) eReaxFF and DFT comparison of equation of state for BZY20 3rd form of Y doping.

In order to study the different aspects of the surface chemistry of BZY20 solid oxide, we built slab models from the BZY20 bulk structure with the 3rd form of Y doping (Fig. 4). We considered two different surface orientations – the (100) and (110). We considered two different terminations for the (100) surface – Ba-O termination and Zr-Y-O termination. From the slab models, we created surface oxygen vacancies at different sites with different concentrations (Fig. 5 and 6). The corresponding eReaxFF and DFT energies for oxygen vacancy formation are summarized in Table 1. Similar to the case of bulk structures, the vacancy formation energies for surfaces are also negative, meaning that the presence of vacancy is not energetically favorable.

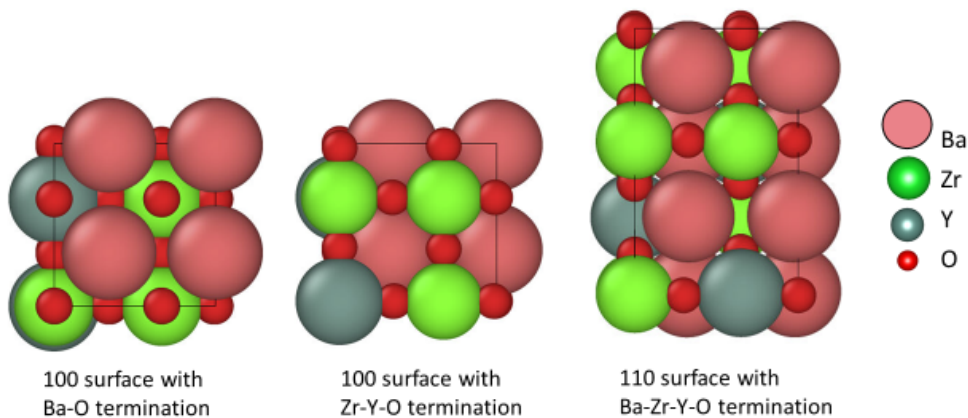

**Fig. 4** | (a) Top view of BZY20 slab models with different surface orientations and surface terminations

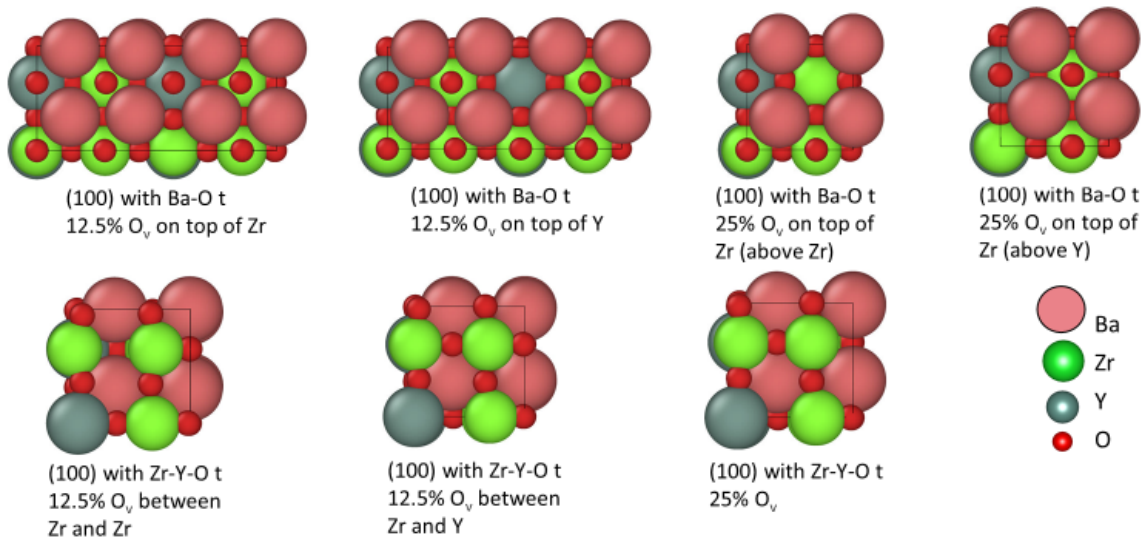

**Fig. 5** | Top view of BZY20 (100) surfaces with different terminations and surface oxygen vacancy ($O_v$) concentrations and sites.

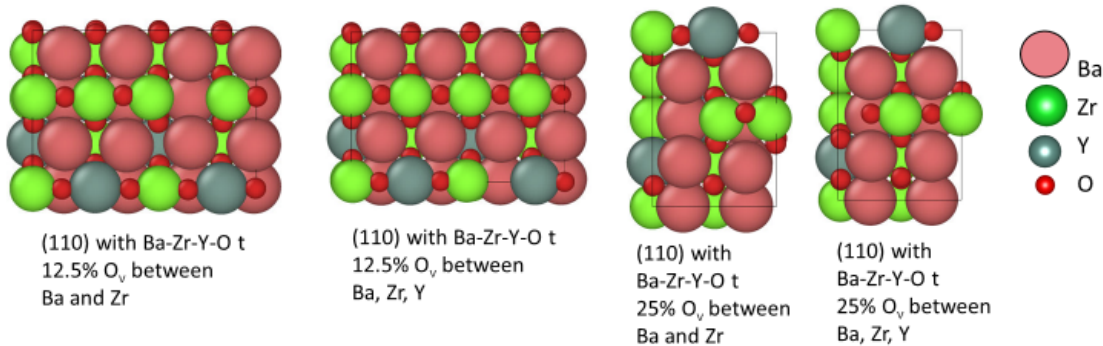

**Fig. 6** | Top view of BZY20 (110) surfaces with Ba-Zr-Y-O termination and different surface oxygen vacancy ($O_v$) concentrations and sites.

**Table 1 | eReaxFF and DFT comparison of BZY20 surfaces with different terminations, surface oxygen vacancy ($O_v$) concentrations and sites**

| Orientation | Termination | $O_v$ concentration at surface | $O_v$ site | $O_v$ formation energy (kcal/mol) eReaxFF | DFT |
|---|---|---|---|---|---|
| (100) | Ba-O | 12.5% | On top of Zr | -141.40 | -236.62 |
| (100) | Ba-O | 12.5% | On top of Y | -92.17 | -99.61 |
| (100) | Ba-O | 25% | On top of Zr (above Zr) | -112.40 | -82.68 |
| (100) | Ba-O | 25% | On top of Zr (above Y) | -103.68 | -115.69 |
| (100) | Zr-Y-O | 12.5% | Between Zr & Zr | -110.96 | -117.74 |
| (100) | Zr-Y-O | 12.5% | Between Zr & Y | -86.65 | -65.69 |
| (100) | Zr-Y-O | 25% | Between Zr & Zr and Zr & Y | -182.42 | -244.01 |
| (110) | Ba-Zr-Y-O | 12.5% | Between Ba & Zr | -119.49 | -98.47 |
| (110) | Ba-Zr-Y-O | 12.5% | Between Ba & Zr & Y | -66.34 | -60.07 |
| (110) | Ba-Zr-Y-O | 25% | Between Ba & Zr | -145.52 | -49.46 |
| (110) | Ba-Zr-Y-O | 25% | Between Ba & Zr & Y | -100.49 | -184.98 |

Finally, we calculated reaction energies for $H_2O$ gas adsorption and $H_2O$ splitting to H and OH where OH would locate at a surface oxygen vacancy site and H would connect to a lattice oxygen. Figure 7 shows the $H_2O$ gas adsorption splitting reaction with energies calculated by eReaxFF and DFT for (100) surface with Zr-Y-O termination with different oxygen vacancy concentrations. Figure 8 shows the $H_2$ gas generation reaction energies calculated by eReaxFF and DFT for the same surface. The eReaxFF calculated energies agree quite well with the DFT energies indicating that the newly developed eReaxFF force field has been trained well with the DFT data.

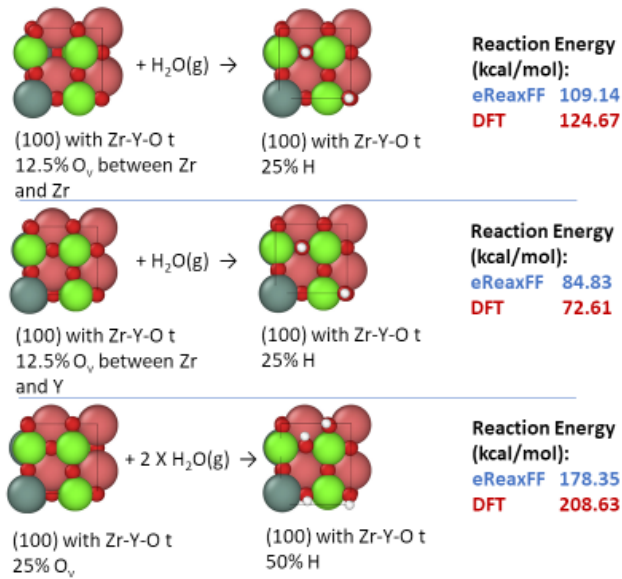

**Fig. 7** | eReaxFF and DFT comparison of H₂O adsorption and splitting reaction energies on BZY20 surfaces.

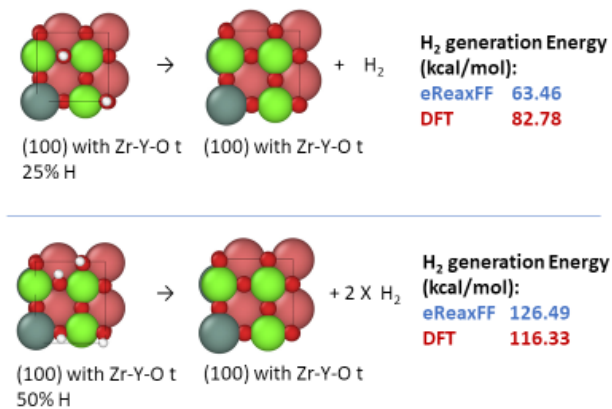

**Fig. 8** | eReaxFF and DFT comparison of H₂ generation energies on BZY20 surfaces.

Based on the above discussion, it can be concluded that our force field is capable of reproducing the DFT data it was trained against quite well. In the next section, molecular dynamics simulations based on this force field will be discussed.

## Results and discussion

Utilizing our newly developed eReaxFF force field, we performed simulations on a BZY20 slab model with Zr Y O terminated surface with 12.5% surface oxygen vacancies. The oxygen vacancies are located between Zr and Y atoms. The mechanism of water adsorption on the oxygen vacancy site and the eventual formation of hydrogen gas is as following:

$H_2O(g) + O_v + O_o \rightarrow OH + OH$  [$O_v$ is an oxygen vacancy site, $O_o$ is a lattice oxygen site]

$2OH \rightarrow 2O_o + 2H$

$2H \rightarrow H_2(g)$

An $H_2O$ molecule adsorbs at the oxygen vacancy site and splits into an H and an OH. H adsorbs on a lattice O, while OH locates at an $O_v$ site. Next, the two OH splits and the two H atoms bind to form $H_2$ gas.

## Bond restraint simulations

In order to determine the binding energy of water molecule and the energy barriers concerning water splitting reactions and hydrogen generation reactions, we performed bond scans using a biasing potential. Figures 9 and 10 shows the different stages and energy barriers of the reactions. From figure 9, it can be seen that water, initially far above the surface (stage 1), finds the oxygen vacancy site and adsorbs with an adsorption energy of 65 kcal/mol (stage 2). Now the water can split into H and OH where OH is situated at the Ov site and H can choose lattice oxygen either between an Y and Zr (Path A) (stage 3b), or between two Zr (Path B) (stage 3a). Path A is more favorable since it has lower energy barrier (7 kcal /mol) than Path B's energy barrier (9 kcal/mol). Also, Path A is more exothermic (-16 kcal/mol) than Path B's exothermicity (-11 kcal/mol). The H dissociation from OH and eventual $H_2$ formation can occur either from stage 2 or state 3b as shown in Figure 10. $H_2$ generation directly from stage 2 is more favorable than from stage 3b but for both cases the energy barriers are extremely high and endothermic indicating that, these reactions might not occur in the solid oxide operating temperature of 800 K. No excess electrons, thus no applied voltage, are present in the solid oxide. It is expected that at applied voltages, the $H_2$ generation barrier might get lowered and $H_2$ generation might occur in the simulations.

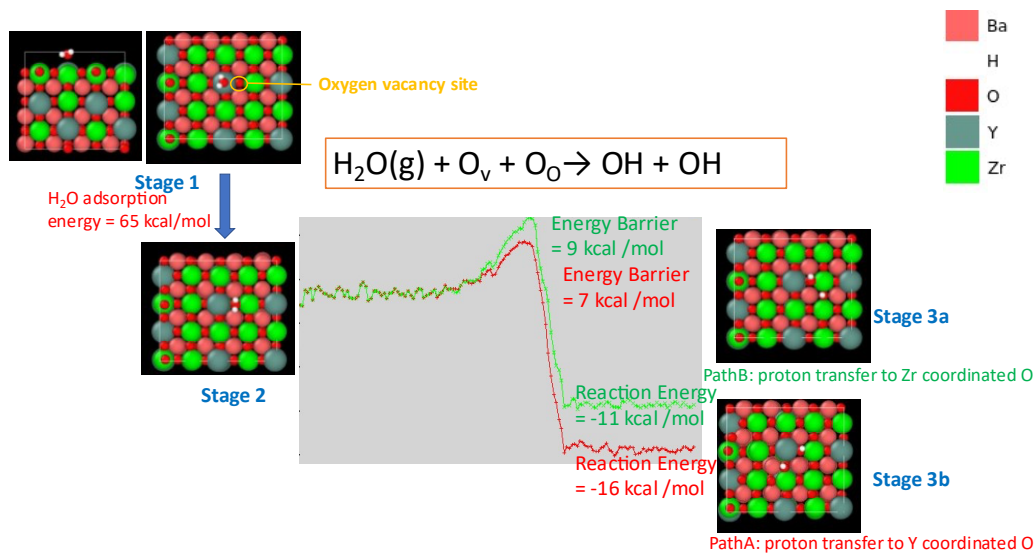

**Fig. 9** | Snapshots of bond restraint simulations and energy barriers and reaction energies associated with water adsorption and water splitting reactions. Color scheme: Ba: pink, H: white, O: red, Y: grey and Zr: green.

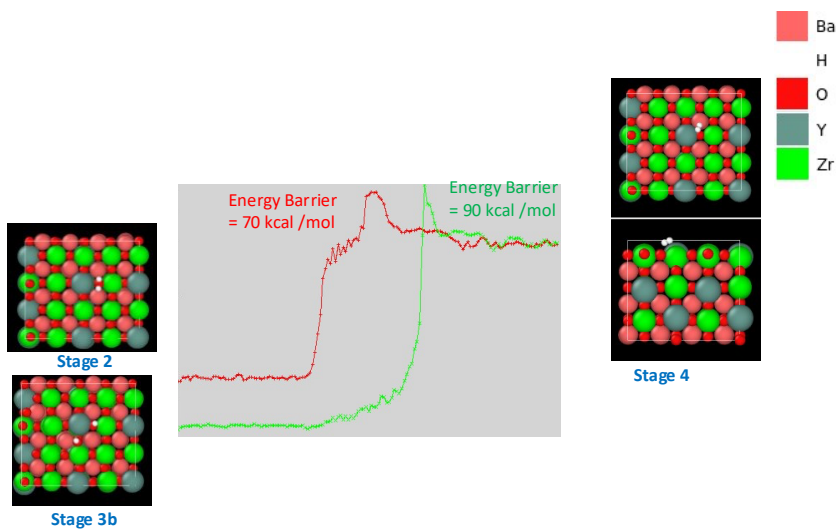

**Fig. 10** | Snapshots of bond restraint simulations and energy barrier associated with hydrogen generation reactions. Color scheme: Ba: pink, H: white, O: red, Y: grey and Zr: green.

On the surface with Ba-O termination, the adsorption energy of water molecule to the oxygen vacancy site ($O_vY$) is -91.1 kcal/mol, which is more stable than the adsorption on the Zr-Y-O termination. The pathways for water splitting reactions are depicted in Fig. 11. Unlike the Zr-Y-O surface, the water splitting reaction on the oxygen vacancy is required to overcome a higher energy barrier (30.4 kcal/mol) with a slightly exothermic reaction energy (-3.8 kcal/mol). Thus, a single water molecule will not easily dissociate on the vacancy site. However, when an extra water molecule assists a proton transfer from the bound water to the surface oxygen (the catalytic assistance of water molecule), it significantly lowers the energy barrier to 13.4 kcal/mol – about a half of the energy barrier for the single water reaction – because of the easier H-transfer through the H-bond network. Water splitting to subsurface layer is endothermic with the energy barrier of 32.8 kcal/mol, indicating that the proton diffusion to subsurface layers rarely occurs at low temperature. Overall, the water splitting reaction on the vacancy site of the Ba-O termination is both thermodynamically and kinetically less favorable than that on Zr-Y-O termination.

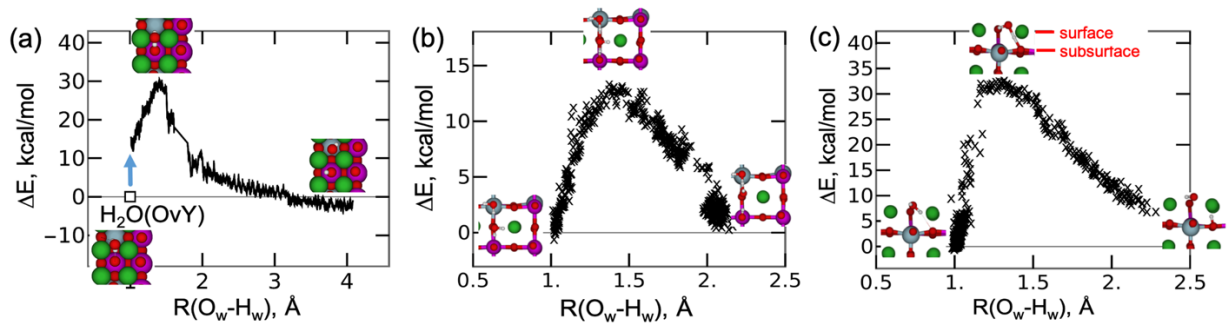

**Fig. 11** Potential energy profiles and structures of water splitting reactions on the surface with Ba-O termination along three pathways - (a) on $O_vY$, (b) on $O_vY$ with water-assistance and (c) on subsurface.

**Molecular dynamics simulations**

In this section we will describe large scale molecular dynamics simulations of steam exposed to BZY20 solid oxide with 12.5% oxygen vacancy between Zr and Y atoms. This gives us 25 vacancy sites on the surface. The density of actual superheated steam at 773 K is 0.00028 g/cc, however, in our simulation we are using is a density of 0.0065 g/cc and a temperature of 1000 K in order to enhance the reaction dynamics over the MD simulation time we chose. We have performed MD NVT simulations for 412 ps. Figure 11a,b show the snapshots at the beginning and end of the simulation, respectively. Throughout the course of the MD simulation, all but one $H_2O$ molecules adsorb at the vacancy sites and the remaining adsorbs at Zr. Although $H_2O$ has a higher adsorption energy for the vacancy site (65 kcal/mol), it is still possible for $H_2O$ to adsorb on Y ($H_2O$ adsorption energy 12 kcal/mol) or Zr ($H_2O$ adsorption energy 37.4 kcal/mol). All adsorbed $H_2O$ molecules split into H and OH and H moves to lattice oxygen sites between Zr and Y, which is a favorable site as we have discussed earlier. Three protons transfer to subsurface oxygens, among them two transfers to the second from the top layer and one transfer to the third layer. This proves the current force field's capability of simulating proton transfer to bulk regions on the solid oxide. This is a zero-voltage simulation, so as expected, no $H_2$ generation is observed due to extremely high energy barrier associated with this reaction. Also continued this simulation with an intermittent flow of steam to observe the saturation of $H_2O$ adsorption on the BZY20 surface. We added 10 $H_2O$ molecules every 400 ps since it takes almost 400 ps for all 10 $H_2O$ molecules to adsorb on the surface. After a total of 110 $H_2O$ molecules are introduced, all but 2 $H_2O$ molecules are adsorbed. All the vacancy sites are filled and the remaining $H_2O$ are either adsorbed on Y or Zr. All adsorbed $H_2O$ splits into H and OH. Proton transfer to 6th layer from the top surface is observed (Fig. 12). Still No $H_2$ generation observed.

    It is expected that, with the implementation of the explicit electron concept in this eReaxFF force field, we would be able to train and simulate non-zero voltage scenarios and expect $H_2$ generation to occur.

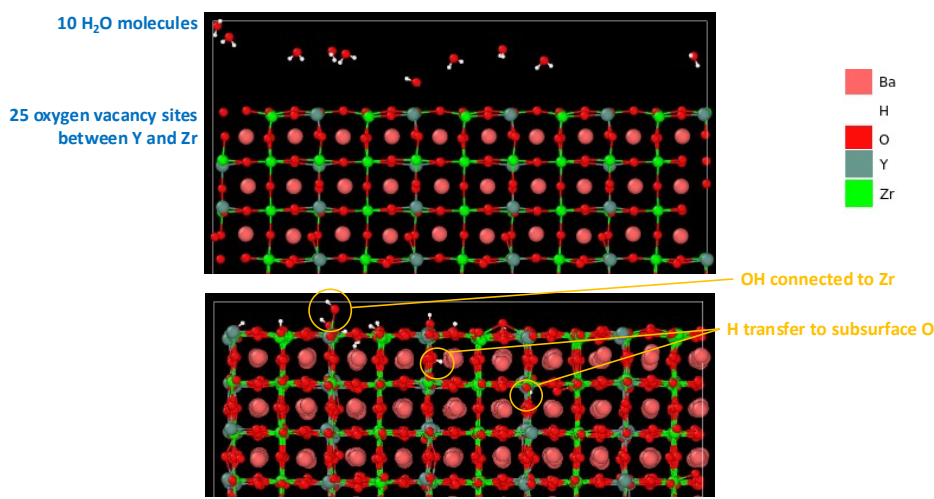

**Fig. 12** | Snapshots of zero-voltage MD simulation of steam adsorption, H$_2$O splitting, proton transfer reactions at (a) $t = 0$ and (b) $t = 412$ ps. Color scheme: Ba: pink, H: white, O: red, Y: grey and Zr: green.

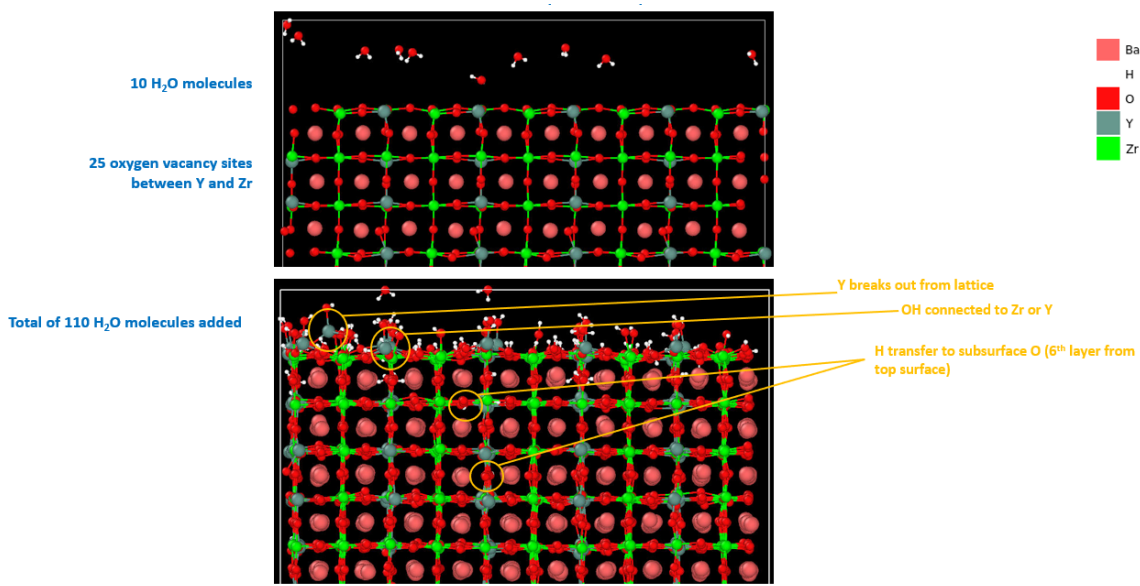

**Fig. 13** | Snapshots of zero-voltage MD simulation of adsorption an intermittent flow of steam, H$_2$O splitting, proton transfer reactions at (a) $t = 0$ and (b) $t = 4.4$ ns. Color scheme: Ba: pink, H: white, O: red, Y: grey and Zr: green.

In contrast, as seen in Fig. 14c, only 41% of water molecules (45 H$_2$O) split into H and OH on the Ba-O terminated surface with 12.5% O$_v$ after 11 cycles of an intermittent flow of steam. Due to the higher energy barrier and the higher reaction energy, the water splitting reaction takes place slowly as we have discussed earlier. The surface coverage of hydroxyl groups is 0.45 ML. The remaining 65 H$_2$O molecules form a thin layer near the surface. Eight protons are transferred to the subsurface or deeper layer oxygens and the deepest proton was observed in the fourth layer from the top surface. No H$_2$ generation is observed. On the defect-free surface, the water splitting reaction is faster than the surface with O-vacancy (Fig. 14c) with the OH coverage of 0.57 ML. By tracking the reactions occurring in the MD simulations, it was observed that unlike the reaction observed in O$_v$-defect surface, a gas phase water molecule directly reacts with the surface oxygen (OY) to transfer a proton and the hydroxyl group of the dissociated water is bound to the neighboring Ba. This hydroxyl group can diffuse fast from one Ba site to the other on the surface (Fig. 15a). The potential energy profile of the water dissociation on the clean surface (Fig. 15) indicates that this reaction is spontaneous with the reaction energy of -5 kcal/mol. As the water-splitting reaction along this pathway is thermodynamically and kinetically more favorable than the pathways described in Fig. 11, this is the predominant mechanism on the Ba-O terminated surface. However, the Zr-Y-O terminated surface is still more active for water dissociation than the Ba-O terminated surface as the surface oxygen density on the surface is higher, indicating that more oxygen sites are accessible to water.

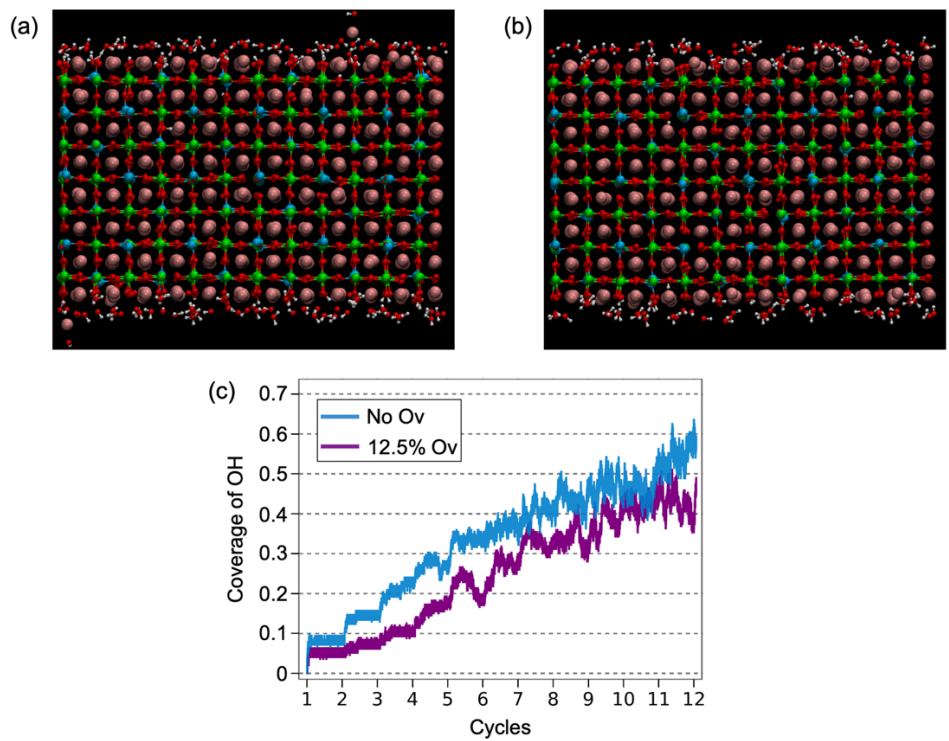

**Fig. 14** Snapshots of zero-voltage MD simulations for water splitting during an intermittent flow of steam with (a) 12.5% $O_v$ and (b) no $O_v$ after 11 cycles. (c) The OH coverage on the surface.

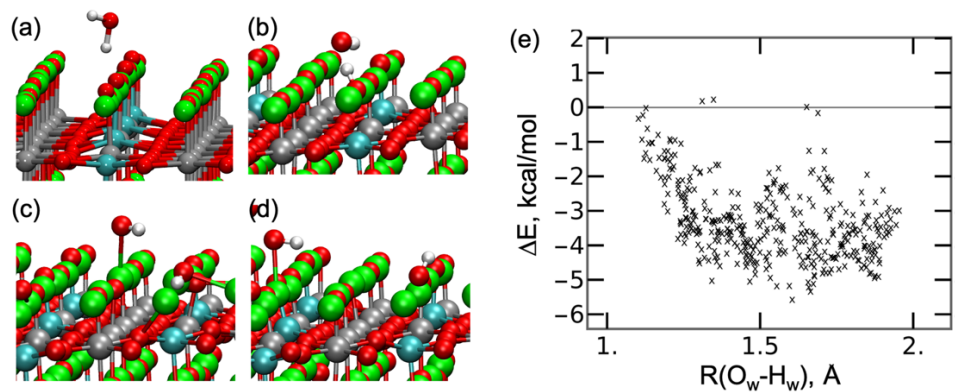

**Fig. 15** (a) Snapshots and (b) potential energy profile of water-splitting reaction on defect-free Ba-O terminated surface.

## Conclusion

We have successfully developed an initial framework for simulating BZY20 solid oxide electrocatalysis. The newly developed eReaxFF force field reproduces the DFT data it was trained against quite well. The zero voltage simulations of steam adsorption on BZY20 surface with oxygen vacancies conclude that hydrogen generation does not occur at operating temperature of 1000 K. Based on our simulation results, we conclude that this force field sets a stage for the introduction of explicit electron concept by further training the eReaxFF with constrained-DFT data in order to simulate electron conductivity, electron leakage and non-zero voltage effects on hydrogen generation. This force field can be extended to include additional dopants to understand the effects on oxygen migrations and hydrogen generation.

## Acknowledgment

This research was supported through the Idaho National Laboratory (INL) Laboratory Directed Research & Development (LDRD) Program under U.S. Department of Energy–Idaho Operations Office (DOE–ID) Contract DE-AC07-05ID14517. Accordingly, the publisher, by accepting the article for publication, acknowledges that the U.S. government retains a nonexclusive, paid-up, irrevocable, worldwide license to publish or reproduce the published form of this manuscript, or allow others to do so, for U.S. government purposes.